# MIMO Transmission with Residual Transmit-RF Impairments

Christoph Studer
Communication Technology Laboratory
ETH Zurich, CH-8092 Zurich, Switzerland
Email: studerc@nari.ee.ethz.ch

Markus Wenk and Andreas Burg
Integrated Systems Laboratory
ETH Zurich, CH-8092 Zurich, Switzerland
Email: {mawenk,apburg}@iis.ee.ethz.ch

*Abstract*—Physical transceiver implementations for multiple-input multiple-output (MIMO) wireless communication systems suffer from transmit-RF (Tx-RF) impairments. In this paper, we study the effect on channel capacity and error-rate performance of residual Tx-RF impairments that defy proper compensation. In particular, we demonstrate that such residual distortions severely degrade the performance of (near-)optimum MIMO detection algorithms. To mitigate this performance loss, we propose an efficient algorithm, which is based on an i.i.d. Gaussian model for the distortion caused by these impairments. In order to validate this model, we provide measurement results based on a 4-stream Tx-RF chain implementation for MIMO orthogonal frequency-division multiplexing (OFDM).

## I. INTRODUCTION

Physical implementations of radio frequency (RF) transmitters suffer from a number of impairments that degrade the quality of the received signal beyond the impact of fading or thermal-noise at the receiver [1]. The most prominent non-idealities are carrier-frequency and sampling-rate offset [2], phase-noise [3], IQ-imbalance [4], and amplifier non-linearities [5]. The impact of some of these impairments can be compensated (partially) at the transmitter through calibration or pre-distortion, or at the receiver using sophisticated compensation algorithms (see [6] for a comprehensive summary on RF impairments and corresponding compensation methods).

However, for all practical scenarios, a certain amount of distortion typically remains unaccounted for. The reasons for the presence of such residual transmit-RF (Tx-RF) impairments are, for example, parameter estimation errors in compensation algorithms (e.g., caused by thermal noise at the receiver), mismatch between the physical RF-chain and the model assumed for impairment compensation, or simply the fact that the compensation algorithms for some impairments are too complex for economic implementation. In multiple-input multiple-output (MIMO) wireless communication systems, distortions resulting from residual Tx-RF impairments that defy proper compensation appear as colored noise at the receiver [3], [7]. Unfortunately, the presence of this noise is still routinely ignored in the development and performance analysis of new transmission schemes and in the design of corresponding MIMO receiver algorithms. Merely the work in [7] reports an impact of Tx-RF impairments on the performance of MIMO detection algorithms, where the *apparent receiver noise* (defined to include thermal noise *and* the residual Tx-RF impairments) is used as reference for the signal-to-noise-ratio (SNR). This definition, however, leads to observations that indicate the performance of linear MIMO detection schemes may improve with the addition of residual Tx-RF impairments.

*Contributions:* In this paper, we extend the conventional MIMO system model with a statistical model for residual Tx-RF impairments. The model is validated using real-world measurement results of a 4-stream Tx-RF chain in a MIMO orthogonal frequency-division multiplexing (OFDM) scenario. In contrary to the work in [7], we define the SNR by the ratio of the received signal power to the thermal noise power only. The use of this definition enables us to study and intuitively explain the impact of residual Tx-RF impairments on the channel capacity and on the performance of well-established MIMO detection algorithms. In particular, we demonstrate that algorithms developed to achieve (near-)optimum performance in absence of residual Tx-RF impairments, suffer from a substantial performance loss even with high-quality RF-chains. These results clearly indicate that the minimum quality requirements for residual Tx-RF impairments defined by modern wireless communication standards (e.g., IEEE 802.11n [8]), are insufficient to justify a careless treatment of the Tx-RF impairment issue for many relevant MIMO communication systems. We therefore propose a method that partially mitigates the impact of residual Tx-RF impairments at the receiver and describe a corresponding efficient algorithm to perform the necessary computations.

*Notation:* Matrices are set in boldface capital letters and vectors in boldface lower-case letters. We write $A_{i,j}$ for the entry in the $i$th row and $j$th column of $\mathbf{A}$. The superscript $^H$ stands for conjugate transpose, $\mathbf{A}^{-1}$ and $\mathbf{A}^\dagger$ for the inverse and Moore-Penrose pseudo-inverse of the matrix $\mathbf{A}$, respectively. $\mathbf{I}_N$ is the $N \times N$ identity matrix. The $\ell^2$-norm of $\mathbf{x}$ is $\|\mathbf{x}\|$. The expectation operator is $\mathbb{E}[\cdot]$ and $\Pr[\mathcal{X}]$ refers to the probability of the event $\mathcal{X}$.

*Outline:* The remainder of the paper is organized as follows. In Sec. II, we introduce the MIMO system model with residual Tx-RF impairments. The impact on channel capacity and performance of MIMO detection algorithms of these impairments is studied in Sec. III and Sec. IV, respectively. A corresponding mitigation method along with the resulting algorithm is provided in Sec. V. Real-world measurement results are provided in Sec. VI. We conclude in Sec. VII.

## II. MIMO SYSTEM MODEL WITH RESIDUAL TX-RF IMPAIRMENTS

We consider a spatial-multiplexing (SM) MIMO wireless communication system with $M_\text{T}$ transmit and $M_\text{R} \geq M_\text{T}$ receive-antennas. Data is mapped to $M_\text{T}$-dimensional transmit symbol-vectors $\mathbf{s} \in \mathcal{O}^{M_\text{T}}$, where $\mathcal{O}$ denotes the set of underlying complex-valued scalar constellation points and $\mathbb{E}\left[\mathbf{s}\mathbf{s}^H\right] = \mathbf{I}_{M_\text{T}}$. Accounting only for the thermal-noise at the receiver leads to the well-known complex baseband input-output relation [9]

$$\mathbf{y} = \mathbf{H}\mathbf{s} + \mathbf{n}_\text{r} \quad (1)$$

where $\mathbf{y}$ is the $M_\text{R}$-dimensional receive-vector and $\mathbf{H}$ denotes the $M_\text{R} \times M_\text{T}$ complex-valued channel matrix. The thermal noise at the receiver is modeled by the $M_\text{R}$-dimensional i.i.d. circularly symmetric complex Gaussian (CSCG) distributed noise vector $\mathbf{n}_\text{r}$ with variance $\sigma_\text{r}^2$ per complex-valued entry, i.e., $\mathbf{n}_\text{r} \sim \mathcal{CN}(0, \sigma_\text{r}^2 \mathbf{I}_{M_\text{R}})$. Throughout the paper, coherent detection is considered, i.e., the receiver knows the realization of the channel matrix $\mathbf{H}$ and the noise variance $\sigma_\text{r}^2$ perfectly, whereas no channel state information is available at the transmitter. The entries of the channel matrix $\mathbf{H}$ are modeled i.i.d. (across space) Rayleigh fading, i.e., $H_{i,j} \sim \mathcal{CN}(0,1)$, $\forall i,j$, and the average SNR per receive antenna is $\text{SNR} = M_\text{T}/\sigma_\text{r}^2$.

### A. Gaussian Model for Residual Tx-RF Impairments

The residual Tx-RF impairments considered in this paper result for example from additive noise, from non-linear distortions in the power amplifier, from inter-carrier interference due to phase noise, or from IQ-imbalance. While only some of these impairments have been reported to match well with Gaussian noise [5], [10], our own measurement results for the MIMO-OFDM case indicate that an i.i.d. additive Gaussian noise model accurately describes the sum of all such residual Tx-RF impairments (see Sec. VI). Furthermore, we assume sufficient decoupling of the Tx-RF chains such that the relevant impairments are statistically independent across transmit antennas. The resulting Gaussian model for the impaired transmitted signal $\tilde{\mathbf{s}}$ is given by

$$\tilde{\mathbf{s}} = \mathbf{s} + \mathbf{n}_\text{t}, \quad \mathbf{n}_\text{t} \sim \mathcal{CN}(0, \sigma_\text{t}^2 \mathbf{I}_{M_\text{T}}) \quad (2)$$

and $\mathbf{n}_\text{t}$ is referred to as Tx-noise in the remainder of the paper.

In the RF-literature, the residual Tx-RF impairments are routinely expressed in terms of the error-vector magnitude (EVM), which is defined as a lump-sum measure that characterizes the quality of the transmitted signal. Formally, the EVM is defined as

$$\text{EVM} = \frac{\mathbb{E}\left[\|\tilde{\mathbf{s}} - \mathbf{s}\|^2\right]}{\mathbb{E}\left[\|\mathbf{s}\|^2\right]}$$

which corresponds to $\text{EVM} = \sigma_\text{t}^2$ for the model in (2). We emphasize that the EVM is typically measured after calibration of the RF-chain and after compensation of those impairments for which robust models and corresponding practical compensation algorithms exist. Typical EVM-values can be obtained from the minimum EVM requirements specified in established standards and from publications of state-of-the-art implementations. For example, IEEE 802.11n [8] specifies -28 dB EVM for 64-QAM modulation and corresponding RF-chain implementations achieve EVMs ranging from -22 dB to -32 dB (depending on the output power), e.g., [11]–[16].

### B. System Model with Residual Tx-RF Impairments

Substituting $\tilde{\mathbf{s}}$ defined in (2) for $\mathbf{s}$ in (1) leads to the following input-output relation of a SM-MIMO system with residual Tx-RF impairments:

$$\mathbf{y} = \mathbf{H}(\mathbf{s} + \mathbf{n}_\text{t}) + \mathbf{n}_\text{r}. \quad (3)$$

In contrast to the thermal noise, the Tx-noise appears as *spatially-colored* noise at the receiver. By combining the two noise sources, i.e., $\tilde{\mathbf{n}} = \mathbf{H}\mathbf{n}_\text{t} + \mathbf{n}_\text{r}$, the input-output relation in (3) can be rewritten in a more compact form as

$$\tilde{\mathbf{y}} = \mathbf{H}\mathbf{s} + \mathbf{K}^{\frac{1}{2}}\mathbf{w}, \quad \mathbf{w} \sim \mathcal{CN}(0, \mathbf{I}_{M_\text{R}}) \quad (4)$$

where the covariance matrix of the aggregate noise at the receiver $\tilde{\mathbf{n}}$ is given by

$$\mathbf{K} = \sigma_\text{t}^2 \mathbf{H}\mathbf{H}^H + \sigma_\text{r}^2 \mathbf{I}_{M_\text{R}} \quad (5)$$

and $\mathbf{K}^{\frac{1}{2}}$ is a square root of $\mathbf{K}$, i.e., $\mathbf{K}^{\frac{1}{2}}\left(\mathbf{K}^{\frac{1}{2}}\right)^H = \mathbf{K}$.

## III. IMPACT OF TX-NOISE ON CHANNEL CAPACITY

Initial insight into the performance limitations arising from residual Tx-RF impairments is obtained from the MIMO channel capacity in the presence of spatially-colored noise based on the covariance matrix $\mathbf{K}$ in (5).

### A. Channel Capacity with Tx-Noise

For a given channel realization $\mathbf{H}$, the capacity of the system described by (4) corresponds to [17]

$$C(\mathbf{H}) = \log_2 \det\left(\mathbf{I}_{M_\text{R}} + \mathbf{K}^{-1}\mathbf{H}\mathbf{H}^H\right) \quad (6)$$

bits per channel use (bpcu). Application of the eigenvalue (EV) decomposition to $\mathbf{H}\mathbf{H}^H = \mathbf{U}\mathbf{\Lambda}\mathbf{U}^H$, where $\mathbf{U}$ is of dimension $M_\text{R} \times M_\text{T}$ with $\mathbf{U}^H\mathbf{U} = \mathbf{I}_{M_\text{T}}$, and $\mathbf{\Lambda}$ is an $M_\text{T} \times M_\text{T}$ diagonal matrix with the EVs $\lambda_i$ ($i = 1, \ldots, M_\text{T}$) on its main diagonal, enables us to rewrite (6) according to

$$C(\mathbf{H}) = \sum_{i=1}^{M_\text{T}} C_i, \quad C_i = \log_2\left(1 + \frac{\lambda_i}{\lambda_i \sigma_\text{t}^2 + \sigma_\text{r}^2}\right). \quad (7)$$

From (7) we can see that a SM-MIMO system that is affected by Tx-noise resolves into the same eigenmodes as the corresponding system without Tx-RF impairments in (1), i.e., the matrix $\mathbf{U}$ diagonalizes $\mathbf{H}\mathbf{H}^H$ as well as $\mathbf{K}^{-1}\mathbf{H}\mathbf{H}^H$.[1]

We furthermore observe that for $\lambda_i \sigma_\text{t}^2 \ll \sigma_\text{r}^2$, e.g., if the EV $\lambda_i$ or the variance of the Tx-noise $\sigma_\text{t}^2$ (i.e., the EVM) is small, we obtain $C_i \approx \log_2\left(1 + \lambda_i/\sigma_\text{r}^2\right)$, which is equivalent to the capacity on the $i$th eigenmode *without* Tx-noise. In contrast, in the case where $\lambda_i \sigma_\text{t}^2 \gg \sigma_\text{r}^2$, e.g., if the EV $\lambda_i$ is large, we have $C_i \approx \log_2\left(1 + 1/\sigma_\text{t}^2\right)$ so that the contribution to channel capacity of the $i$th eigenmode is limited by the Tx-noise. This observation implies that strong eigenmodes are more severely affected by Tx-noise than eigenmodes associated with small EVs $\lambda_i$. In addition, for $\textsf{SNR} \to \infty$ we have

$$C_\text{lim} = \lim_{\textsf{SNR}\to\infty} C(\mathbf{H}) = M_\text{T} \log_2\left(1 + 1/\sigma_\text{t}^2\right)$$

which implies that the channel capacity in the presence of Tx-noise is upper-bounded by $C_\text{lim}$.

---

[1]In the presence of CSI at the transmitter, we can furthermore see from (7) that optimal water-filling leads to the same power allocation in the presence of Tx-noise as it would be the case for (1).

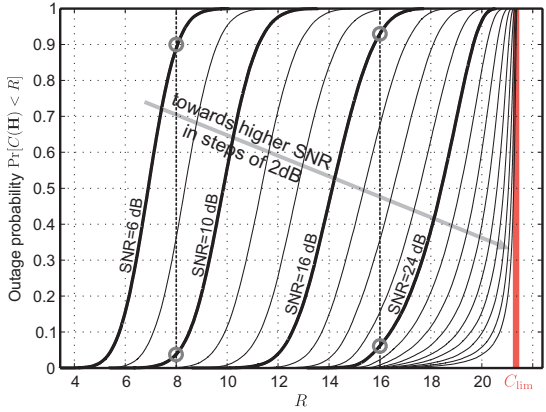

Fig. 1. CDFs of the ergodic channel capacity $C(\mathbf{H})$ in a $M_\text{T} = M_\text{R} = 4$ MIMO system with Tx-noise and -16 dB EVM.

### B. Numerical Channel-Capacity Results

In order to illustrate the above described capacity behavior, we consider the outage probability given by the cumulative distribution function (CDF) of the channel capacity $p_\text{out} = \Pr\left[C(\mathbf{H}) < R\right]$, where $R$ refers to the transmission rate. Fig. 1 provides corresponding simulation results with $M_\text{T} = M_\text{R} = 4$ and an EVM of -16 dB for different SNR-values. The maximum capacity $C_\text{lim}$ for this EVM corresponds to approximately 21 bpcu.

In the low-SNR regime (i.e., for low rates), the CDFs for different values of SNR show the same channel capacity characteristics. However, as the rate $R$ approaches $C_\text{lim}$, the CDFs become skewed and the upper tail disappears, while the lower tail (corresponding to channel realizations with small EVs) remains. Since the outage probability $p_\text{out}$ is a lower bound on the frame error-rate (FER) performance [18], we can also gain insight into the FER behavior as a function of SNR by fixing the rate $R$. In Fig. 1 we observe that the SNR-increase required for the same reduction in terms of FER is much smaller for low rates compared to that for rates close to $C_\text{lim}$. In other words, the FER-slope decreases as the rate increases. This behavior is illustrated in Fig. 1 by the SNR-increase required to go from 90% to 5% outage probability for 8 bpcu and for 16 bpcu. For the lower rate, a 4 dB SNR increase is sufficient, while for the higher rate, an increase of at least 8 dB is required to achieve the desired improvement in terms of FER. In summary, prohibitively high SNRs are required in the presence of Tx-noise in order to reliably achieve transmission rates that are close to the capacity limit $C_\text{lim}$.

## IV. IMPACT OF TX-NOISE ON PERFORMANCE OF MIMO DETECTION ALGORITHMS

Channel capacity considerations provide insight into the theoretical performance limits without consideration of the actual MIMO detection algorithm. For practical purposes, we are, however, particularly interested in the impact of residual Tx-RF impairments on the performance of MIMO detection algorithms that were designed under the idealistic assumption of impairment-free transmitted signals.

For the investigation of this issue, we rely on numerical simulations and consider a coded MIMO system with $M_\text{T} = M_\text{R} = 4$ using 16-QAM modulation and a rate-1/2 convolutional code (generator polynomial [$133_\text{o}$ $171_\text{o}$], constraint length 7, and random interleaving). We assume a block-fading scenario in which 64 symbol-vectors in a frame (corresponding to 1024 coded bits) experience the same channel realization $\mathbf{H}$. Coding is performed over all 64 transmit vectors and the Viterbi algorithm is used for decoding. In the simulation, no phase-noise is assumed and hence, phase-noise tracking (PT) is not performed. The Tx-RF impairments are modeled as described in Sec. II. We use an EVM of -30 dB, which is already at the upper limit of state-of-the-art RF-chain implementations (see, e.g., [16]) and is, for example, 14 dB better than the minimum EVM required by the IEEE 802.11n standard [8] for the transmission rate used in our simulations.

### A. Brief Review of MIMO Detection Algorithms

We consider the following prominent MIMO detection algorithms: zero-forcing (ZF) detection, maximum-likelihood (ML) detection, and (soft-output) max-log a posteriori probability (APP) detection.

*1) ZF detection:* This linear MIMO detection algorithm estimates the transmitted vector by pre-multiplying the received symbol-vector $\mathbf{y}$ with the pseudo-inverse $\mathbf{H}^\dagger$ of the channel matrix, followed by entry-wise quantization of $\mathbf{H}^\dagger \mathbf{y}$ to the nearest constellation point in $\mathcal{O}$, i.e., $\hat{\mathbf{s}}^\text{ZF} = \text{Q}\left(\mathbf{H}^\dagger \mathbf{y}\right)$, where $\text{Q}(\cdot)$ denotes the entry-wise quantization operator [9].

*2) ML detection:* This algorithm follows a Bayesian approach to estimate the transmitted symbol-vectors, which requires knowledge about the statistics of the noise vector $\mathbf{n}$. The ML detection rule for the idealistic i.i.d. CSCG noise model in (1) corresponds to [9]

$$\hat{\mathbf{s}}^\text{ML} = \arg\min_{\mathbf{s}\in\mathcal{O}^{M_\text{T}}} \|\mathbf{y} - \mathbf{H}\mathbf{s}\|^2. \quad (8)$$

*3) Max-log APP detection:* This MIMO detection algorithm computes soft-outputs in the form of log-likelihood ratios (LLRs) $L(b_i)$ for all bits $b_i$ comprising a symbol-vector. Similar to ML detection, the corresponding detection rule is usually formulated for the idealistic input-output relation (1) and uses the max-log approximation to obtain [19]

$$L(b_i) \approx \frac{1}{\sigma_\text{r}^2}\left(\min_{\mathbf{s}\in\mathcal{X}_i^1} \|\mathbf{y} - \mathbf{H}\mathbf{s}\|^2 - \min_{\mathbf{s}\in\mathcal{X}_i^0} \|\mathbf{y} - \mathbf{H}\mathbf{s}\|^2\right) \quad (9)$$

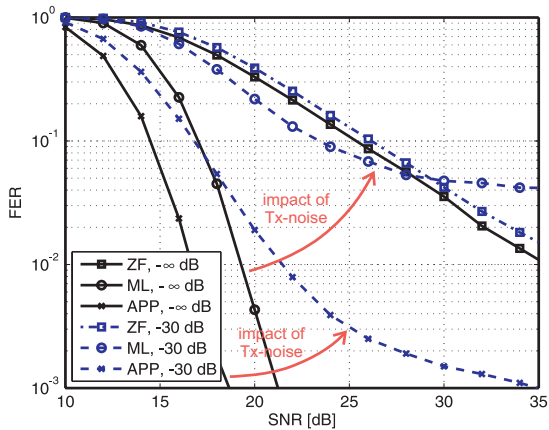

Fig. 2. Impact of -30 dB EVM to ZF, ML, and max-log APP detection.

where $\mathcal{X}_i^1$ and $\mathcal{X}_i^0$ denote the sets of all possible transmit symbol-vectors for which the $i$th bit is equal to one or zero, respectively. Note that computation of (9) essentially corresponds to solving multiple constrained ML detection problems (8), e.g., [20].

*B. Impact of Tx-Noise on Performance*

In the following discussion, we use the MIMO detection algorithms described above and replace the impairment-free received vector $\mathbf{y}$ by $\tilde{\mathbf{y}}$ from (4) which is affected by Tx-noise. Fig. 2 shows the corresponding performance in an ideal system ($-\infty$ dB EVM) and in the presence of $-30$ dB EVM. Without residual Tx-RF impairments, ZF detection achieves by far the worst performance of all three MIMO detection algorithms and ML detection is outperformed by the soft-output max-log APP detector. However, in the presence of weak Tx-noise, it can clearly be observed that both ML and max-log APP detection suddenly suffer from a *substantial* performance loss, whereas the performance of ZF detection is only slightly affected.[2] Note that the performance of ML detection is even worse than that of ZF detection in the high-SNR regime.

*1) ZF detection:* An intuitive explanation for the minor performance degradation of the ZF detector in the presence of residual Tx-RF impairments can be obtained by considering the output of the equalization stage

$$\mathbf{H}^\dagger \tilde{\mathbf{y}} = \mathbf{s} + \mathbf{n}_\mathrm{t} + \mathbf{H}^\dagger \mathbf{n}_\mathrm{r}. \quad (10)$$

The Tx-noise $\mathbf{n}_\mathrm{t}$ is spatially white and independent of $\mathbf{H}$. However, the thermal-noise $\bar{\mathbf{n}} = \mathbf{H}^\dagger \mathbf{n}_\mathrm{r}$ after equalization is colored and suffers from noise-enhancement. Hence, the equalized thermal-noise $\bar{\mathbf{n}}$ dominates over a large SNR-range and the error induced by $\mathbf{n}_\mathrm{t}$ is small for reasonably low EVMs.

*2) ML detection:* In contrast to linear MIMO detection schemes, ML detection relies heavily on the assumption that the noise at the receiver is i.i.d. CSCG, which is *not* the case in presence of Tx-RF impairments. The simulation results in Fig. 2 clearly show that a mismatch between the actual system model with Tx-RF impairments in (4) and the idealistic model in (1) has a detrimental impact on the performance of ML detection.

---

[2]Numerical simulations have shown that the impact on the performance of linear hard-output minimum mean-square error (MMSE) detection is similar to that of hard-output ZF detection.

*3) Max-log APP detection:* The behavior of the max-log APP detector is similar to that of the ML detector, since both algorithms rely on a Bayesian estimation approach and suffer from the same mismatch of the noise statistics. However, the impact of the Tx-noise on the max-log APP detector is less pronounced (compared to that of ML detection) since max-log APP detection reaches a given FER already at a lower SNR (compared to ML detection), where the contribution of the residual Tx-RF impairments to the overall noise is less significant.

## V. MITIGATION OF RESIDUAL TX-RF IMPAIRMENTS

So far, we have argued that the ML and max-log APP detectors both suffer significantly from residual Tx-RF impairments. In order mitigate this performance loss, MIMO detection algorithms can either be formulated for the impaired system model (4) or the i.i.d. CSCG noise distribution must be restored prior to MIMO detection. In order to preserve the structure of well-established MIMO detection algorithms, we shall focus on the second approach in the following.

*A. Tx-Noise Whitening*

Noise whitening offers a straightforward solution to restore the i.i.d. CSCG distribution contained in the received signal of (4). To this end, we define a noise-whitening filter $\mathbf{W}$ based on (5) as

$$\mathbf{W} = \sigma_\mathrm{r} \mathbf{K}^{-\frac{1}{2}} = \sigma_\mathrm{r} \left( \sigma_\mathrm{t}^2 \mathbf{H}\mathbf{H}^H + \sigma_\mathrm{r}^2 \mathbf{I}_{M_\mathrm{R}} \right)^{-\frac{1}{2}} \quad (11)$$

assuming that the EVM (and hence $\sigma_\mathrm{t}^2$) is known at the receiver. Application of (11) to the received vector in (4) yields

$$\mathbf{W}\tilde{\mathbf{y}} = \tilde{\mathbf{H}}\mathbf{s} + \tilde{\mathbf{n}}_\mathrm{r}$$

where $\tilde{\mathbf{H}} = \mathbf{W}\mathbf{H}$ is the effective channel matrix and the resulting additive noise vector $\tilde{\mathbf{n}}_\mathrm{r} \sim \mathcal{CN}(0, \sigma_\mathrm{r}^2 \mathbf{I}_{M_\mathrm{R}})$ has the same statistics as $\mathbf{n}_\mathrm{r}$ in (1). Hence, the MIMO detection algorithms initially developed for i.i.d. CSCG noise can now be applied to $\mathbf{W}\tilde{\mathbf{y}}$ and $\tilde{\mathbf{H}}$ instead of $\tilde{\mathbf{y}}$ and $\mathbf{H}$.

*B. Numerical Performance Results for Tx-Noise Whitening*

Fig. 3 shows the impact of Tx-noise whitening on the performance of ZF, ML, and max-log APP detection in the presence of $-30$ dB EVM. As expected, the performance of the ZF detector remains unaltered since the whitening filter $\mathbf{W}$ cancels out when substituting $\tilde{\mathbf{H}}$ for $\mathbf{H}$ and $\mathbf{W}\tilde{\mathbf{y}}$ for $\tilde{\mathbf{y}}$ in (10). However, Tx-noise whitening yields a substantial improvement in terms of performance for both ML and max-log APP detection compared to that of the mismatched MIMO detectors used in Fig. 2. A clear advantage of ML and max-log APP detection over linear MIMO detection schemes is again visible. In addition, max-log APP performance now outperforms ML detection by approximately 6 dB SNR (for 1% FER), which implies that soft-output detection in combination with coding additionally reduces detrimental effects caused by Tx-RF impairments (see Sec. IV-B3). Nevertheless, a performance penalty remains for all MIMO detection algorithms compared to the scenario without any residual Tx-RF impairments.

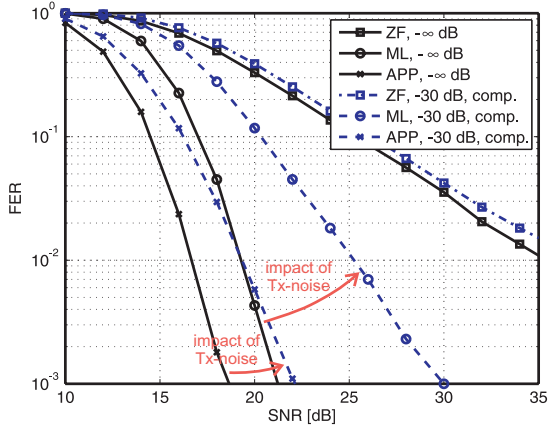

Fig. 3. Performance impact of Tx-noise whitening (denoted by comp.) to ZF, ML, and max-log APP detection impaired by the i.i.d. Gaussian model for residual Tx-RF impairments.

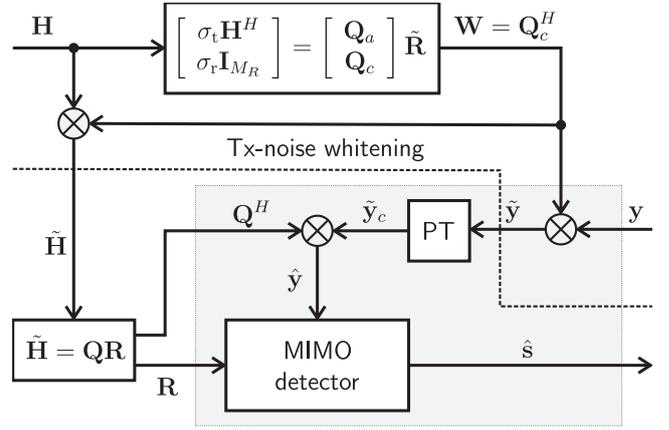

Fig. 4. Data-dependency graph of QRD-based Tx-noise whitening. Note that only those tasks within the gray area need to be performed at symbol-rate. Phase-noise tracking (PT) takes place after Tx-noise whitening.

### C. QRD-Based Tx-Noise Whitening Algorithm

Most of the additional computations required by the proposed noise-whitening algorithm are confined to the preprocessing of the channel matrices, which must only be performed when the channel-state changes. Nevertheless, straightforward computation of $\mathbf{K}^{-\frac{1}{2}}$ in (11) requires explicit computation of the covariance matrix $\mathbf{K}$ in (5) followed by an EV-decomposition or a Cholesky decomposition of $\mathbf{K}$ combined with matrix inversion. Both approaches suffer from high computational complexity and require considerable arithmetic precision, which renders corresponding practical (fixed-point) implementations difficult and potentially uneconomic.

*1) Efficient computation of the whitening filter:* A computationally less complex and numerically more stable approach for computing the whitening filter $\mathbf{W}$ in (11) is obtained from the economy-size QR-decomposition (QRD) of

$$\begin{bmatrix} \sigma_{\mathrm{t}}\mathbf{H}^H \\ \sigma_{\mathrm{r}}\mathbf{I}_{M_{\mathrm{R}}} \end{bmatrix} = \begin{bmatrix} \mathbf{Q}_a \\ \mathbf{Q}_c \end{bmatrix} \tilde{\mathbf{R}} \qquad (12)$$

where $\tilde{\mathbf{R}}$ is an $M_{\mathrm{R}} \times M_{\mathrm{R}}$ upper-triangular matrix with non-negative real-valued entries on the main diagonal and $\mathbf{Q}_a$ and $\mathbf{Q}_c$ are of dimension $M_{\mathrm{T}} \times M_{\mathrm{R}}$ and $M_{\mathrm{R}} \times M_{\mathrm{R}}$, respectively. With (12), the whitening filter in (11) is immediately given by $\mathbf{W} = \mathbf{Q}_c^H$, since $\tilde{\mathbf{R}}^H \tilde{\mathbf{R}} = \mathbf{K}$ in (5) and $\mathbf{Q}_c = \sigma_{\mathrm{r}}\tilde{\mathbf{R}}^{-1}$. We emphasize that computation of the whitening filter according to (12) avoids explicit computation of the covariance matrix $\mathbf{K}$, relaxing the precision requirements of the subsequent steps, and does *not* require an EV-decomposition or a Cholesky decomposition followed by a matrix inversion.

*2) Whitening for QRD-based MIMO detectors:* Since most of the available (near-)optimum MIMO detection algorithms are based on the QRD of the channel matrix[3] $\mathbf{H}$, we show the integration of our QRD-based Tx-noise whitening algorithm into a QRD-based MIMO detector in Fig. 4.

The preprocessing task of the channel matrix $\mathbf{H}$ starts with the computation of the Tx-noise whitening filter $\mathbf{W}$ based on (12). Next, the effective channel matrix $\tilde{\mathbf{H}} = \mathbf{W}\mathbf{H}$ is calculated. Then, the QRD of the effective channel matrix $\tilde{\mathbf{H}} = \mathbf{Q}\mathbf{R}$ is obtained. The symbol-rate tasks (i.e., tasks that need to be performed for each received vector) correspond to i) application of the noise whitening filter $\mathbf{W}$ to the received vectors $\tilde{\mathbf{y}} = \mathbf{W}\mathbf{y}$, ii) compensation of phase-noise in the whitened receive vector $\tilde{\mathbf{y}}$ using the PT unit, and iii) left-multiplication of the compensated vector $\tilde{\mathbf{y}}_c$ by the matrix $\mathbf{Q}^H$, i.e., $\hat{\mathbf{y}} = \mathbf{Q}^H \tilde{\mathbf{y}}_c$. The MIMO detector finally produces estimates of the transmitted data symbols $\hat{\mathbf{s}}$ (or LLRs) based on $\hat{\mathbf{y}}$ and the upper-triangular matrix $\mathbf{R}$.

*3) Computational complexity:* In the above configuration, the complexity required by the Tx-noise-whitening algorithm is mainly in the preprocessing stage. In particular, the additional preprocessing computations correspond to an economy-size QRD of an $(M_{\mathrm{R}} + M_{\mathrm{T}}) \times M_{\mathrm{R}}$ matrix and one matrix-by-matrix multiplication. Note that the proposed algorithm offers potential for reusing of arithmetic components that are already present in preprocessing units of many state-of-the-art MIMO receivers (see, e.g., [21]). The additional complexity required at symbol-rate only corresponds to pre-multiplication of the received vectors $\mathbf{y}$ with the Tx-noise whitening matrix $\mathbf{W}$.

## VI. REAL-WORLD MEASUREMENT RESULTS

In order to validate the i.i.d. Gaussian Tx-noise model for residual transmit-RF impairments and the impact of Tx-noise-whitening as a means to mitigate the associated performance degradation, real-world measurements on a 4-stream RF-chain implementation have been performed. We characterize the statistics of real-world residual Tx-RF impairments and show

---

[3]The long list of QRD-based detection algorithms includes for example, sphere decoding or any of the other tree-search based MIMO detection algorithms (see, e.g., [19], [20]).

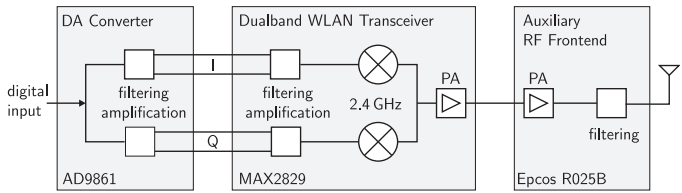

Fig. 5. Transmit RF-chain used for proof-of-concept and for measurements to validate the impaired system model and Tx-noise whitening.

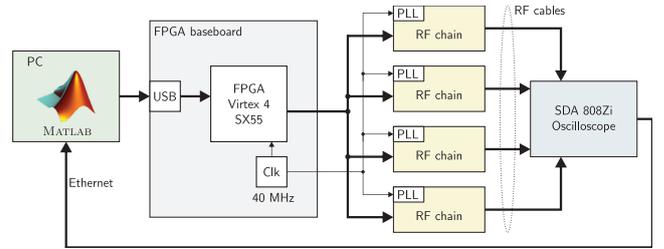

Fig. 6. The measurement setup consists of a PC running MATLAB, an FPGA base-board, four (calibrated) RF-chains, and an SDA 808Zi oscilloscope.

corresponding frame error-rate (FER) performance results in a MIMO-OFDM setting.

### A. RF-Chain Implementation and Measurement Setup

We next summarize the main components of the RF-chain implementation and provide details on the measurement setup.

*1) RF-chain implementation:* The RF-chain is identical to the one used in the real-time multi-user MIMO-OFDM testbed presented in [22]. The corresponding block diagram is shown in Fig. 5. The implementation consists of a two-channel 10-bit digital-to-analog converter (DAC) chip (Analog Devices, AD9861) with 80 MSPS, a dual-band direct-conversion RF transceiver chip (Maxim Integrated Products, MAX2829) specifically designed for OFDM 802.11 WLAN applications, and an auxiliary RF front-end module (Epcos, R025B) for filtering and amplification. Facilities to calibrate carrier leakage and IQ-imbalance are available, i.e., the DAC provides gain-control and offset registers that allow for adjustment of the output current gain and the current offset for the I- and Q-path independently. All measurements are performed using a *calibrated* RF-chain.

*2) Measurement setup:* The measurement setup is illustrated in Fig. 6. A PC using MATLAB is used for measurement control and signal processing. The FPGA baseboard stores the data to be transmitted. Four calibrated RF-chains are used and each RF-chain contains its own phase-locked loop (PLL). Note that this configuration is sub-optimal and renders phase-noise tracking difficult. To measure the impact of residual Tx-RF impairments only, a high-performance oscilloscope (LeCroy, SDA 808Zi) is responsible for RF data acquisition at 2.4 GHz.

The measurements are performed as follows. Each transmitted MIMO-OFDM frame is generated in MATLAB and transferred to the FPGA. Then, the entire frame is transmitted through the four RF-chains and recorded (simultaneously) on the oscilloscope. The recorded frame is then transferred back to MATLAB, where mixing, down-sampling, synchronization (i.e., in terms of frequency and timing), and OFDM baseband-processing are performed (see Fig. 8 for more details).

### B. Characterization of Residual Tx-RF Impairments

We first validate the i.i.d. Gaussian Tx-noise assumption for modeling of residual Tx-RF impairments. In order to isolate the residual Tx-RF impairments from the transmitted MIMO-OFDM frame, the (known) transmitted signal is subtracted from the measured samples. The EVM of the residual Tx-RF impairments was determined according to the IEEE 802.11n standard [8] and is $-28$ dB.

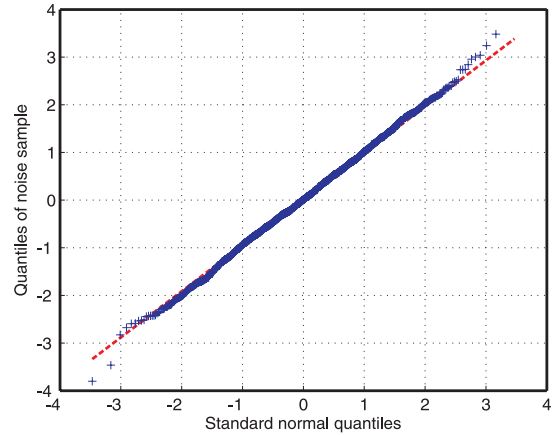

Fig. 7. Quantile-quantile plot comparing a real-valued standard normal distribution with measured samples of our RF-chain implementation.

*1) Statistics of residual Tx-RF impairments:* Fig. 7 compares a standard normal distribution with the measured samples from our Tx-RF chain implementation using a quantile-quantile plot [23]. The samples used in Fig. 7 correspond to the real-part of the measured residual Tx-RF impairments resulting from a single RF-chain and from one frame consisting of 40 OFDM symbols. However, our measurements consistently show similar results for the imaginary part and for the remaining RF-chains. To account for crosstalk between RF-chains, four streams have been transmitted simultaneously. For the sake of clarity of exposition, the resulting data stream is normalized to unit variance. As it can be observed in Fig. 7, the assumption that the residual Tx-RF impairments are Gaussian distributed is accurate. We emphasize that this observation is valid for the employed RF-chain and the considered OFDM scenario, but not necessarily valid for other RF-chain implementations or other transmission schemes.

*2) Inter- and intra-stream correlation:* In order to validate the assumption that the Tx-noise can be modeled as i.i.d. (among streams, real and imaginary parts) random variables, we correlated the real and imaginary parts of the acquired residual Tx-RF impairments. In addition, the empirical correlation between the signals from the four RF-chains was computed. We neither observed strong correlation between the four spatial streams nor between the I- and Q-paths, which is mainly a result of calibrating the RF-chain components previous to performing the measurements. In particular, the strongest off-diagonal component of the empirical covariance matrix was less than 19.5 dB below that of the diagonal entries. Hence, the i.i.d. assumption holds for the employed RF-chain implementation and the considered MIMO-OFDM scenario.

*C. FER Performance Measurements*

In order to study the impact of real-world residual Tx-RF impairments on the performance of MIMO-OFDM systems, FER performance measurements were carried out. In addition, these FER results are compared to simulation results using the Tx-noise model proposed in Sec. II-A.

*1) FER measurement setup:* FER measurements are carried out in a $M_T = M_R = 4$ MIMO-OFDM scenario by transmitting frames that consist of four short preambles, two long preambles, four MIMO-training symbols, a signal field and 32 OFDM data symbols. Each OFDM data symbol contains 64 tones, where 48 are data sub-carriers and four are used for PT (refer to [22], [24] for more details on the employed MIMO-OFDM frame structure). The rate-1/2 convolutional code described in Sec. IV is used in combination with 16-QAM modulation, resulting in 3072 coded bytes (corresponding to 1536 information bytes) per frame.

Fig. 8 illustrates the receiver that is used for the FER-performance comparison between the Tx-noise model introduced in Sec. II-A and real-world measurements of residual Tx-RF impairments. For the Tx-noise model simulation, $-32\,\mathrm{dB}$ i.i.d. distributed Gaussian transmit noise was added to the frames generated in MATLAB. This amount of Tx-noise results in an EVM of about $-28\,\mathrm{dB}$, since measuring the EVM according to IEEE 802.11n [8] includes errors induced by channel estimation and synchronization. In addition, the effective channel of the Tx-RF chain (i.e., corresponding to up-sampling filters etc.) and phase-noise was modeled (denoted by "model for RF-chain" in Fig. 8). In order to obtain the FER performance curves, the following steps of the simulation are swept for different receive-noise levels (according to the desired SNR) and 1000 frames were transmitted over 1000 different TGn type C channel realizations [25]. The receiver performs channel estimation, applies the Tx-noise whitening algorithm described in Sec. V-C, and compensates phase-noise *after* noise whitening in the PT unit (based on pilot tones). The MIMO-processing unit includes the QRD of the whitened channel matrix $\tilde{\mathbf{H}}$ and MIMO detection. Finally, the frame is decoded and the FER is computed.

*2) FER performance results:* Fig. 9 shows the measured and simulated FER for ZF detection and ML detection. The measurement results show that Tx-noise whitening significantly improves the FER in the presence of (real-world) residual Tx-RF impairments.[4] We furthermore observe that the performance associated with simulated Tx-RF impairments (corresponding to the dotted lines) match well to the FER-performance resulting from measured data. We therefore conclude that the proposed Tx-noise model accurately characterizes residual Tx-RF impairments in MIMO-OFDM systems.

[4]The measurement results and the corresponding simulations differ from the simulation curves provided in Fig. 2 and Fig. 3. For the simulations in Sec. V-A, a block-fading channel model is used and each frame consist of only one OFDM symbol. In addition, perfect channel estimation, synchronization, and no phase noise is assumed there.

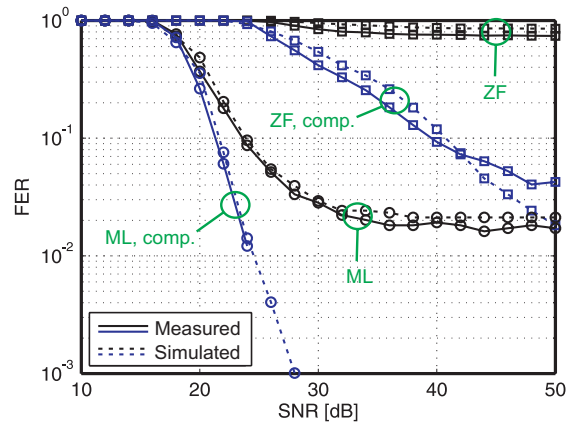

Fig. 9. Measured FER for ZF and ML detection and the impact of noise whitening (denoted by comp.). The dashed lines show the corresponding simulation results based on the Tx-noise model.

## VII. CONCLUSION

Residual transmit (Tx) radio-frequency (RF) impairments are often ignored in the literature on multiple-input multiple-output (MIMO) wireless communication systems. In this paper, we demonstrated that residual Tx-RF impairments do have a detrimental impact on MIMO channel capacity and even small error-vector magnitudes (EVMs) severely affect the performance of (near-)optimum MIMO detection algorithms and should therefore be taken into account in the corresponding performance analysis. In particular, Bayesian MIMO detection algorithms with excellent performance for systems without residual Tx-RF impairments (such as, e.g., maximum-likelihood or soft-output max-log a posteriori probability detectors), suffer significantly from such impairments.

We proposed Tx-noise whitening as an efficient means to mitigate this performance loss. The corresponding Tx-noise whitening filter can be computed efficiently with an economy-size QR-decomposition (QRD), which approximately doubles the complexity of the preprocessing task, but has no impact on the complexity of QRD-based MIMO detection algorithms itself. These properties render the proposed mitigation method suitable for implementation in practical systems.

Real-world measurements have been conducted in order to characterize the statistics and the impact of residual Tx-RF impairments on the performance of MIMO orthogonal frequency-division multiplexing (OFDM) systems. We observed that the proposed i.i.d. Gaussian Tx-noise model accurately reflects real-world residual Tx-RF impairments. We finally demonstrated that Tx-noise whitening considerably mitigates the performance loss associated with residual Tx-RF impairments in practical systems.

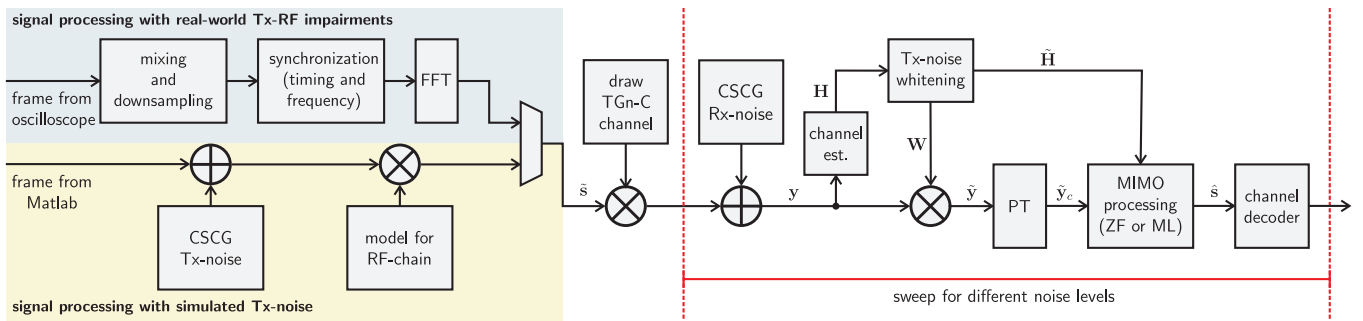

Fig. 8. Signal-processing of the MIMO-OFDM receiver used to obtain measured and simulated FER performance results.

ACKNOWLEDGMENTS

The authors would like to thank H. Bölcskei, M. Borgmann, D. Cescato, U. Schuster, D. Seethaler, and P. Tejera for fruitful discussions on residual transmit-RF impairments and corresponding mitigation methods. This work was partially supported by the STREP project No. IST-026905 (MASCOT) within the Sixth Framework Programme (FP6) of the European Commission. Furthermore, financial support from the Swiss National Science Foundation under the project number PP002-119052 is gratefully acknowledged.